\documentclass[acmsmall]{acmart}
\usepackage{listings}

\usepackage{xcolor}

\lstdefinelanguage{json}{
    morestring=[b]",
    morecomment=[l]{//},
    morekeywords={true,false,null},
    stringstyle=\color{red},
    keywordstyle=\color{blue},
    commentstyle=\color{gray},
    literate={:}{{\textcolor{orange}{:}}}{1},
}

\usepackage{tcolorbox}
\tcbuselibrary{listings}

\tcbset{
  mylisting/.style={
    colback=white,
    colframe=black,
    boxrule=0.5pt,
    arc=2pt,
    outer arc=2pt,
    width=0.48\textwidth,
    height=6cm, 
    listing only,
    listing options={
      language=C,
      basicstyle=\ttfamily\footnotesize,
      numbers=none,
      showstringspaces=false
    }
  }
}

\lstdefinestyle{ide}{
  backgroundcolor=\color{gray!10},   
  basicstyle=\ttfamily\footnotesize, 
  breaklines=true,                   
  frame=single,                      
  framerule=0pt,                     
  rulecolor=\color{gray!40},         
  columns=fullflexible,              
  keepspaces=true,                   
  showstringspaces=false,            
}
\usepackage{graphicx}

\AtBeginDocument{%
  }

\setcopyright{acmlicensed}
\copyrightyear{2025}
\acmYear{2025}
\acmDOI{XXXXXXX.XXXXXXX}

\acmJournal{JACM}
\acmVolume{37}
\acmNumber{4}
\acmArticle{111}
\acmMonth{8}

\begin{document}

\title{Can LLMs Recover Program Semantics? A Systematic Evaluation with Symbolic Execution}

\author{Rong Feng}
\email{rjf5768@psu.edu}
\author{Suman Saha}
\email{sumsaha@psu.edu}
\affiliation{%
  \institution{The Pennsylvania State University}
  \city{University Park}
  \state{PA}
  \country{USA}
}








\renewcommand{\shortauthors}{Feng et al.}

\begin{abstract}

Obfuscation poses a persistent challenge for software engineering tasks such as program comprehension, maintenance, testing, and vulnerability detection. While compiler optimizations and third-party code often introduce transformations that obscure program intent, existing analysis tools and large language models (LLMs) struggle to recover the original semantics. In this work, we investigate whether LLMs, when fine-tuned with symbolic execution artifacts, can effectively deobfuscate programs and restore analyzability. We construct a benchmark by applying four widely studied transformations—control-flow flattening, opaque predicates, arithmetic encoding, and branch encoding—across diverse C programs from TUM Obfuscation Benchmarks, the LLVM test suite, and algorithmic repositories. We then compare three state-of-the-art LLMs under two training configurations: baseline fine-tuning on obfuscated/original code pairs, and enhanced fine-tuning with additional KLEE artifacts such as SMT constraints, path statistics, and test cases. Our evaluation examines syntactic correctness (compilation success), semantic fidelity (behavioral equivalence under symbolic execution), and code quality (readability and structure). Results show that GPT-4.1-mini achieves the strongest deobfuscation overall, and that incorporating KLEE artifacts consistently improves semantic preservation and compilation success across models. These findings highlight deobfuscation as a broader software engineering concern, demonstrating that combining LLMs with symbolic execution can strengthen automated testing, static analysis, and program comprehension in the presence of obfuscation.

\end{abstract}



\received{20 February 2007}
\received[revised]{12 March 2009}
\received[accepted]{5 June 2009}

\maketitle

\section{Introduction}

Obfuscation is widely used in modern software systems. Developers rely on it to protect intellectual property, while malware authors use it to conceal malicious intent. Regardless of motivation, obfuscation complicates key software engineering activities such as program comprehension, debugging, maintenance, and vulnerability detection~\cite{schrittwieser2016protecting,xu2017secure}. Transformations like control-flow flattening~\cite{laszlo2009obfuscating,blazy2016formal}, opaque predicates~\cite{xu2016generalized,tung2020heuristic}, and virtualization~\cite{salwan2018symbolic,liang2017deobfuscation} are intentionally designed to confuse both human analysts and automated tools. This makes deobfuscation---recovering analyzable and semantically faithful code---an essential challenge for the software engineering community.

Existing techniques, however, have clear limitations. Symbolic execution has been a cornerstone of program analysis for decades~\cite{baldoni2018survey,duraibi2019survey}. Tools like \textit{KLEE} can generate high-coverage test inputs and uncover deep bugs~\cite{cadar2008klee}, but symbolic execution suffers from path explosion and quickly becomes impractical when applied to heavily obfuscated programs~\cite{sebastian2017predicting,ollivier2019obfuscation}. Machine learning approaches have also been explored. For example, Banerjee et al.\ and He et al.\ recovered variable names and debug information from stripped binaries~\cite{banerjee2021variable,he2018debin}, while Su et al.\ applied deep learning to Android deobfuscation~\cite{su2018obfuscation}. Although these approaches achieve partial success, they often fail to generalize across different obfuscation strategies.

Large language models (LLMs) have recently emerged as powerful tools for program analysis. They have been used for variable and function name recovery~\cite{xu2025unleashing,jiang2025beyond}, decompilation~\cite{tan2024llm4decompile,hu2024degpt}, and even code readability improvement~\cite{zou2025d}. Some studies have specifically tested LLMs on deobfuscation benchmarks, such as \textit{JsDeObsBench}~\cite{chen2025jsdeobsbench} and broader multi-transformation datasets~\cite{beste2025exploring}. These results show that LLMs can simplify and partially restore obfuscated code, but they are prone to generating code that does not compile or preserve semantics~\cite{tkachenko2025deconstructing}. In short, neither symbolic execution nor LLMs alone can reliably deobfuscate complex programs.

This paper explores a new direction: combining symbolic execution artifacts with LLM fine-tuning. Our intuition is that symbolic execution provides hard semantic evidence (e.g., SMT constraints, test inputs, execution paths), while LLMs bring flexible reasoning and code synthesis. By integrating these two sources of strength, we obtain models that are more reliable in producing compilable and semantically faithful code. To the best of our knowledge, no prior work has systematically combined symbolic execution with LLMs for deobfuscation.

We make the following contributions:
\begin{enumerate}
    \item Reframing deobfuscation as a software engineering challenge central to comprehension, testing, and maintenance, rather than solely a security concern.
    \item Constructing a dataset that integrates programs from the TUM repository, the LLVM test suite, and a public Algorithms repository, transformed with four widely studied obfuscations.
    \item Demonstrating that incorporating \textit{KLEE}-generated artifacts into fine-tuning improves LLM performance across syntax, semantics, and readability.
    \item Providing empirical evidence across three representative LLMs and four obfuscation types, showing that our hybrid approach consistently outperforms baseline fine-tuning.
\end{enumerate}

To guide our evaluation, we design research questions that assess syntax, semantics, readability, and transformation sensitivity. We introduce these formally in Section~4.

\section{Background}

\subsection{Code Obfuscation}
Code obfuscation means changing a program so that it is harder to read or analyze, while still producing the same results. In practice, developers use it to protect proprietary software, while attackers rely on it to hide malware~\cite{schrittwieser2016protecting,xu2017secure}. A few transformations are especially common. Control-flow flattening replaces a clear branching structure with a dispatcher loop, which makes the paths look uniform and confusing~\cite{laszlo2009obfuscating,blazy2016formal}. Opaque predicates are conditions that always evaluate the same way, but appear complicated enough to mislead both humans and tools~\cite{xu2016generalized,tung2020heuristic}. Arithmetic encoding takes simple expressions and replaces them with much longer algebraic forms, while branch encoding rewrites conditional logic in indirect ways. On their own, these tricks are simple, but when used together they can dramatically increase the cost of analysis.

\subsection{Symbolic Execution}
Symbolic execution is a program analysis technique that runs programs with symbolic inputs instead of concrete values. Rather than producing one output, it creates logical constraints that describe many possible paths~\cite{baldoni2018survey,duraibi2019survey}. This makes it a powerful tool for automatically generating test cases or finding deep bugs. \textit{KLEE}, for example, showed how symbolic execution could uncover real errors in widely used system programs~\cite{cadar2008klee}. Still, symbolic execution has well-known limits. One of the biggest is the path explosion problem: every new branch multiplies the number of paths to explore, and the number grows exponentially~\cite{sebastian2017predicting,ollivier2019obfuscation}. Solvers also struggle with complex arithmetic and system-level interactions. These weaknesses explain why symbolic execution alone is often ineffective when applied to obfuscated programs, which are designed to create exactly such difficult cases.

\subsection{Large Language Models in Program Analysis}
Large language models (LLMs) have recently been applied to a range of program analysis tasks, from variable name recovery to full decompilation~\cite{banerjee2021variable,xu2025unleashing,jiang2025beyond,tan2024llm4decompile}. Their strength comes from recognizing statistical patterns across large code corpora and generating code that looks natural to developers. This flexibility makes them attractive for reverse engineering, where strict static rules often fail. But LLMs also have well-documented shortcomings. Sometimes they output code that compiles but changes behavior, or code that preserves behavior but no longer compiles cleanly. In other cases, they hallucinate, inventing functions or data structures that were never there in the first place~\cite{tkachenko2025deconstructing}. For software engineering tasks that demand reliability, these issues remain a major obstacle.

\subsection{Why Deobfuscation Is Hard}
Deobfuscation combines all of these difficulties. Obfuscation deliberately hides structure; symbolic execution gets lost in the resulting explosion of paths; and LLMs may produce code that looks plausible but is semantically wrong. When several obfuscations are stacked, the challenge only grows. For software engineers, this is not an abstract issue: maintainers often need to work with obfuscated third-party code, and analysts need to test or audit software that has been deliberately transformed. At present, no single approach---neither symbolic execution nor LLMs---can reliably restore obfuscated programs to a clean, analyzable form. Overcoming this challenge requires new methods that bring together the precision of formal analysis with the adaptability of machine learning.

\section{Approach}

\begin{figure}[!htbp]
  \centering
  \includegraphics[width=\textwidth]{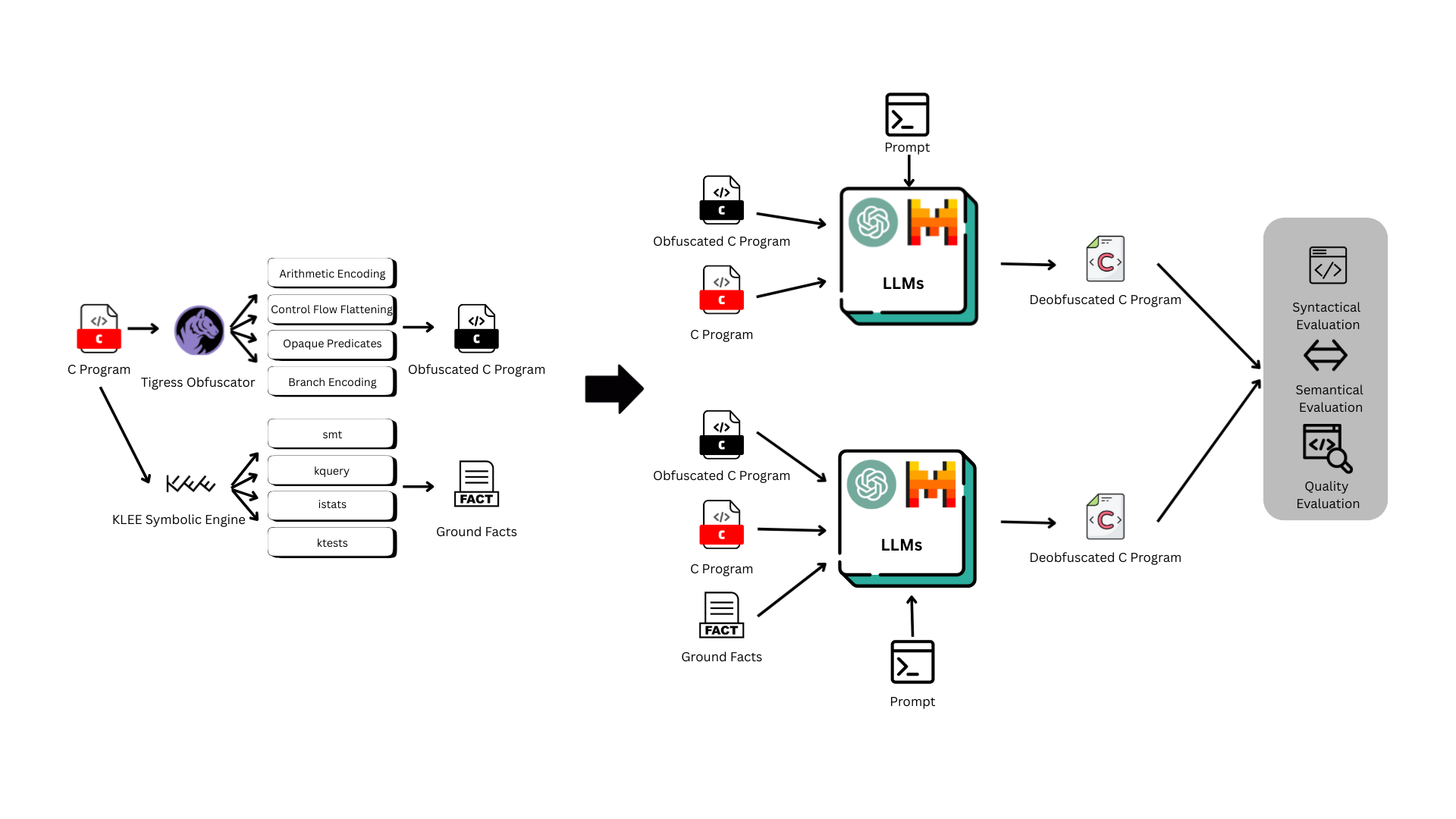}
  \caption{Overview of our approach. Source programs are obfuscated with four transformations, while KLEE generates semantic artifacts from the original code. These inputs guide LLM fine-tuning, producing deobfuscated programs that are evaluated for syntax, semantics, and quality.}
  \label{fig:workflow}
\end{figure}

Our approach combines program obfuscation, symbolic execution, and large language model (LLM) fine-tuning in a unified workflow (Figure~\ref{fig:workflow}). We begin with source programs drawn from three repositories and apply four widely studied obfuscation techniques to create challenging inputs. In parallel, we use \textit{KLEE} on the original programs to produce symbolic artifacts, such as constraints, path statistics, and test cases, which serve as semantic ground truth. These artifacts are then paired with the obfuscated code and used to fine-tune LLMs under two configurations: a baseline setting that relies only on code, and an enhanced setting that incorporates symbolic artifacts. At inference time, the models are prompted with obfuscated programs and expected to generate deobfuscated outputs. We then evaluate these outputs using syntactic, semantic, and quality measures. The following subsections describe each step of the workflow in detail: dataset construction, obfuscation transformations, symbolic execution artifacts, model training, and data formatting.

\subsection{Dataset}
\label{subsec:dataset}

We constructed our dataset from three complementary sources to capture programs of varying size and complexity. The first source was the TUM repository, which provides small but well-structured C programs widely used in software engineering benchmarks. The second source was the LLVM test suite, which contains medium-sized programs covering diverse computational patterns. Finally, we included programs from a public Algorithms repository, consisting of implementations of common algorithms such as sorting, searching, and dynamic programming. Together, these sources provided a broad set of code fragments suitable for evaluating deobfuscation.

Since obfuscation can drastically increase program size and symbolic reasoning costs, we applied two filtering steps. First, we excluded programs that led to path explosion in \textit{KLEE}. In practice, this removed a significant number of programs where symbolic execution generated an intractable number of execution paths. Second, we filtered programs based on context size, discarding those that produced inputs exceeding the maximum token window of our target LLMs. After filtering, we retained a balanced set of programs across the three repositories. To evaluate transformation-specific effects, we applied four obfuscation techniques---control-flow flattening (CFF), opaque predicates (OP), arithmetic encoding (AE), and branch encoding (BE)---to each program. Table~\ref{tab:dataset} summarizes the final dataset, showing the number of programs per obfuscation type along with approximate lines-of-code (LoC) ranges.

\begin{table}[h!]
\centering
\caption{Dataset summary after filtering. Each program was obfuscated with all four transformations (AE, CFF, OP, BE), producing a total of 680 variants.}
\label{tab:dataset}
\begin{tabular}{lcc}
\hline
\textbf{Source} & \textbf{\#Programs} & \textbf{LoC Range (Each File)} \\
\hline
TUM Repository   & 99 & 9--109\\
LLVM Test Suite  & 39 & 7--829\\
Algorithms Repo  & 32 & 73--425  \\
\hline
\textbf{Total}   & 170 & 7--829 \\
\hline
\end{tabular}
\end{table}

\subsection{Obfuscation Transformations}

To create challenging inputs for our study, we applied four widely studied obfuscation transformations: control-flow flattening (CFF), opaque predicates (OP), arithmetic encoding (AE), and branch encoding (BE). These transformations were chosen because they distort program structure at both the control-flow and data-flow levels, and they are commonly implemented in tools such as Obfuscator-LLVM. For each base program, we generated four obfuscated variants, one per transformation. Below we illustrate each transformation with a small example.

\subsubsection{Control-Flow Flattening (CFF)}
CFF replaces structured branching with a dispatcher loop and a switch statement, obscuring the natural flow of execution.

\begin{minipage}{0.5\textwidth}
\begin{lstlisting}[language=C, caption={Before}]
int f(int x) {
    if (x > 0) {
        return x + 1;
    } else {
        return x - 1;
    }
}
\end{lstlisting}
\end{minipage}
\hfill
\begin{minipage}{0.5\textwidth}
\begin{lstlisting}[language=C, caption={After (CFF)}]
int f(int x) {
    int state = 0;
    while (1) {
        switch(state) {
            case 0: state = 
                (x > 0) ? 
                1 : 2; break;
            case 1: 
                return x + 1;
            case 2: 
                return x - 1;
        }
    }
}
\end{lstlisting}
\end{minipage}

\subsubsection{Opaque Predicates (OP)}
OP introduces conditions that always evaluate to the same outcome but appear complex, misleading analysis tools.

\begin{minipage}{0.5\textwidth}
\begin{lstlisting}[language=C, caption={Before}]
int g(int y) {
    if (y % 2 == 0) {
        return y / 2;
    } else {
        return 3 * y + 1;
    }
}
\end{lstlisting}
\end{minipage}
\hfill
\begin{minipage}{0.5\textwidth}
\begin{lstlisting}[language=C, caption={After (OP)}]
int g(int y) {
    if (((y*y - y*y) + 1) > 0) 
        { // always true
        if (y % 2 == 0)
            return y / 2;
        else
            return 3 * y + 1;
    } else {
        return 0; 
            // unreachable
    }
}
\end{lstlisting}
\end{minipage}

\subsubsection{Arithmetic Encoding (AE)}
AE rewrites simple arithmetic expressions into longer but equivalent forms.

\begin{minipage}{0.5\textwidth}
\begin{lstlisting}[language=C, caption={Before}]
int h(int a) {
    return a + 1;
}
\end{lstlisting}
\end{minipage}
\hfill
\begin{minipage}{0.5\textwidth}
\begin{lstlisting}[language=C, caption={After (AE)}]
int h(int a) {
    return (a * 4 + 4) / 4; 
        // equivalent to a + 1
}
\end{lstlisting}
\end{minipage}

\subsubsection{Branch Encoding (BE)}
BE hides branch conditions by encoding them through bit manipulations or tables.

\begin{minipage}{0.5\textwidth}
\begin{lstlisting}[language=C, caption={Before}]
int k(int z) {
    if (z > 0) return 1;
    else return -1;
}
\end{lstlisting}
\end{minipage}
\hfill
\begin{minipage}{0.45\textwidth}
\begin{lstlisting}[language=C, caption={After (BE)}]
int k(int z) {
    int cond = (z >> 31) & 1; 
            // encodes sign
    if (cond == 0) return 1;
    else return -1;
}
\end{lstlisting}
\end{minipage}

\noindent
Together, these transformations significantly reduce readability and frustrate conventional program analysis. By applying them systematically, we created four distinct obfuscated variants of each base program, resulting in a diverse and challenging dataset for evaluation.

\subsection{KLEE Artifacts}

To provide semantic ground truth for deobfuscation, we applied KLEE to the original (unobfuscated) programs. KLEE produces multiple artifacts that capture both symbolic reasoning and concrete execution behavior. These artifacts were paired with the obfuscated programs and incorporated into model fine-tuning.

\begin{itemize}
    \item \textit{SMT2 constraints.} These encode path conditions in SMT-LIB format and provide solver-ready assertions describing program behavior. For example:
\begin{center}
\begin{lstlisting}[language=C]
(assert (> x 0))
\end{lstlisting}    
\end{center}

    \item \textit{KQuery expressions.} KLEE’s internal intermediate representation of symbolic constraints. KQuery is more compact than SMT2 and reflects the structural form of the constraints.

    \item \textit{Istats reports.} Summary statistics about symbolic execution, including the number of explored paths, solver calls, and timeouts. These values characterize the complexity of each program.

    \item \textit{KTest files.} Concrete inputs generated to satisfy path constraints. For example:
\begin{lstlisting}[language=C]
object 0: name: "x", data: 5
\end{lstlisting}
    Such test cases act as executable evidence that link symbolic constraints to real program behavior.
\end{itemize}
These artifacts capture both symbolic path constraints and concrete behaviors, providing semantic ground truth that the enhanced model can exploit during fine-tuning.

\subsection{Model Configuration}
\label{subsec:model-configuration}
For each of the four obfuscation transformation techniques--Control Flow Flattening (CFF), Opaque Predicates (OP), Arithmetic Encoding (AE), and Branch Encoding(BE)--and for each LLM model that is being evaluated, we have designed two different training configurations. This approach was intended to examine how well symbolic execution artifacts help LLM for deobfuscation.  

The first configuration is the Baseline Model (code pairing between the original program and the obfuscated program). It follows a straightforward supervised fine-tuning method, in which several previous works have been done. During fine-tuning, the model is presented with the obfuscated code as input and the original code as output. This code pairing forces the model to learn and understand how the obfuscation transformation changes the structure and syntactic patterns of a program.   

The second configuration is the Enhanced Model (in addition to code pairing, we are also providing KLEE Ground Truth). This builds on the same input-output pairing but adds additional input information with symbolic execution artifacts generated by KLEE as ground truth. The expected output remains the same as the original program, but now the model has more information about the execution semantics and the path feasibility information. This model is designed to test whether LLMs can use KLEE Ground Facts to improve deobfuscation quality, particularly when facing challenging transformation techniques such as opaque predicates and deep nested control flow structures.  

This setup allows us to examine the result with fair comparison: any improvement or better statistics can be attributed to the inclusion of symbolic execution artifacts, rather than other influences.
\subsection{Training Data Format}
\label{subsec:training-data-format}
All data is transformed into JSONL one-liner format to ensure compatibility with the majority fine-tuning APIs. Every message is structured like a prompt-response pair. From the figure below, we can see that we use system message to set up the role and the task the LLM is asked to do; we use the user message to provide obfuscated programs (and KLEE artifacts if applicable); we use the assistant message to provide the answer we want to see from LLM during the fine-tuning process. During testing, the assistant message will be filled in by the fine-tuned LLM. Below are two sample JSONL one-liners that we have been using for fine-tuning:  
\subsection*{Example Prompts for Model Training}

\noindent\textbf{Baseline Model (without KLEE):}
\begin{lstlisting}[style=ide, language=json]
{"messages": [
  {"role": "system", "content": "You are an expert deobfuscation assistant. Output only valid C code."},
  {"role": "user", "content": "Obfuscated Code:\n<obfuscated snippet>"},
  {"role": "assistant", "content": "<original clean program>"}
]}
\end{lstlisting}

\noindent\textbf{Enhanced Model (with KLEE Artifacts):}
\begin{lstlisting}[style=ide, language=json]
{"messages": [
  {"role": "system", "content": "You are an expert deobfuscation assistant. Use the provided artifacts when available."},
  {"role": "user", "content": "Obfuscated Code:\n<obfuscated snippet>\n\nKLEE Artifacts:\nSMT2: <path constraints>\nKQuery: <query representation>\niStats: <execution stats>\nKTest: <test cases>"},
  {"role": "assistant", "content": "<original clean program>"}
]}
\end{lstlisting}

By maintaining the same target output for both model configurations, we ensured that the only experiment variable is whether KLEE artifacts are presented in the input or not. This design provides a controlled framework to examine the effect of KLEE on the deobfuscation ability of LLM.
\subsection{Models Used}
\label{subsec:models-used}
For our fine-tuning and evaluation experiments, we selected three state-of-the-art large language models (LLMs) that represent different specialties in terms of capacity, specialization, and context length. By comparing these models with identical deobfuscation tasks, we aimed to understand the trade-off between open-source code-focused models, lightweight models, and closed-source frontier models.  

The first model is GPT-4.1-mini. It is a frontier model that is accessed through the API. GPT-4.1-mini provides an insanely large context length window of up to 1,047,576 tokens. This number far exceeds most existing large language models. This characteristic makes it a great fit for tasks that involve large training and output files. Prior benchmarks have also demonstrated that GPT-4.1-mini consistently delivers strong results in code generation and reasoning tasks. This model has also been involved with many prior work related to obfuscation or LLM fine-tuning, which makes it an easy choice to use to pick during the planning phase.  

The second model is Ministral-8B-latest, a recently released open-source model with 8 billion parameters and a 128k token context length window. Unlike GPT-4.1-mini, Ministal represents efficiency and deployability, making it attractive for research scenarios where local inference or reduced hardware requirements are important. While this model has a smaller context window compared to GPT-4.1-mini, Ministral has been observed to deliver outputs that are readable and stylistically consistent. Including Ministral allows us to evaluate how a lightweight model will perform on code deobfuscation compared to GPT.  

The third mode is Codestral-latest, known as the code-specialized model in the open-source community. With a 256k token context length window, Codestral has a balance between capacity and specialization. For our study, Codestral serves as an alternative strong model for comparison with GPT-4.1-mini. 

\begin{table}[htbp]
\centering
\caption{Model Context Lengths}
\label{tab:model_context}
\resizebox{0.6\textwidth}{!}{
\begin{tabular}{|l|l|r|}
\hline
\textbf{Model} & \textbf{Model ID} & \textbf{Context Length} \\
\hline
GPT & GPT-4.1-mini-2025-04-14 & 1,047,576 \\
Ministral & Ministral-8B-latest & 128,000 \\
Codestral & Codestral-latest & 256,000 \\
\hline
\end{tabular}
}
\end{table}

\section{Experimental Setup}
\label{sec:experimental-setup}

\subsection{Research Questions}
\label{subsec:rqs}

We structure our empirical study around the following research questions:

\begin{description}
  \item[RQ1 (Syntax):] How often do LLMs produce compilable deobfuscated code across different obfuscation types, and how does KLEE guidance impact compilation success?
  \item[RQ2 (Semantics):] To what extent do deobfuscated programs preserve behavior relative to the original programs, and how much does KLEE improve semantic fidelity?
  \item[RQ3 (Readability/Quality):] Do LLMs improve readability (naming, structural simplicity, documentation) over obfuscated inputs, and how do models differ in quality?
  \item[RQ4 (Comparison):] How do fine-tuning (vs.\ base models) and with-KLEE (vs.\ no-KLEE) affect syntax, semantics, and quality?
  \item[RQ5 (Transformation Sensitivity):] Which obfuscation transformations (AE, BE, CFF, OP) are easiest or hardest for LLMs, and does KLEE shift those boundaries?
\end{description}

\subsection{Tasks and Metrics}
\label{subsec:metrics}

We evaluate model performance along three axes aligned with RQ1--RQ3: syntactic correctness, semantic preservation, and readability/quality.

\subsubsection{Syntactic Correctness (RQ1)}
\label{subsubsec:syntax-metric}

For each deobfuscated output, we ask a binary question: \emph{does the code compile to LLVM bytecode (.bc) or not?}
Each file receives a score of 100 for successful compilation and 0 otherwise. We recorded the success rate per model x training condition x transformation, which is a total of 24 entries. To ensure fair comparison, we only score files that appear in both the with-KLEE and without-KLEE conditions.

\subsubsection{Semantic Preservation (RQ2)}
\label{subsubsec:semantic-metric}

To assess semantic preservation, we compare KLEE-generated test counts between (i) the original unobfuscated program and (ii) the LLM-deobfuscated program.
Intuitively, if the deobfuscated program preserves behavior, KLEE should be able to explore a similar number of paths and generate a similar number of test cases.
For each pair, we compute a normalized score in $[0, 100]$ based on the difference in test counts and report the mean by transformation, model, and training condition.

\subsubsection{Readability and Quality (RQ3)}
\label{subsubsec:quality-metric}

We score readability/quality on a 0--100 scale using a static analysis over all deobfuscated C programs, focusing on three aspects:

\begin{itemize}
  \item \textbf{Naming.} Ratio of meaningful identifiers, with penalties for likely obfuscated tokens such as single-letter names and long vowel-less strings.
  \item \textbf{Structural simplicity.} Cyclomatic complexity, maximum nesting depth, and average function length.
  \item \textbf{Documentation.} Presence of comments and basic docstrings where applicable.
\end{itemize}

These signals are combined into a single quality score per file; we again report mean per transformation, model, and condition.

\subsection{Dataset and Difficulty Context}
\label{subsec:dataset-eval}

The dataset construction, sources, and filtering criteria are described in Section~\ref{subsec:dataset}.
In brief, we select 170 C programs from three repositories (TUM, LLVM test suite, and an Algorithms repository) and apply four obfuscation transformations (AE, BE, CFF, OP), yielding 680 obfuscated variants (Table~\ref{tab:dataset}).

For evaluation, we use the same filtered set of programs and additionally compute statistics on the obfuscated inputs and deobfuscated outputs (tables omitted for space).
Across the corpus, obfuscated programs are substantially larger than their originals; opaque predicates (OP) are especially challenging, often producing very large outputs with significantly inflated lines of code.
Even after deobfuscation, OP remains the hardest regime: outputs are shorter but still relatively large, indicating partial simplification but persistent complexity.
These statistics contextualize the compilation and semantic scores reported in Section~\ref{sec:results}.

\subsection{Model Variants and Training Conditions}
\label{subsec:models-conditions}

As outlined in Section~\ref{subsec:model-configuration}, for each obfuscation transformation (AE, BE, CFF, OP) and each LLM, we consider two training configurations:

\begin{itemize}
  \item \textbf{Baseline (code-only).} The model is fine-tuned in a supervised way where the input is the obfuscated program and the target is the original program, measuring how far LLMs can go by learning the mapping between obfuscated and clean code without explicit semantic guidance.
  \item \textbf{Enhanced (code + KLEE).} The model receives the same input--output pair, but the user message additionally includes KLEE artifacts: SMT2 and KQuery path constraints, \texttt{istats} execution summaries, and \texttt{ktest} concrete inputs. The target output remains the original program, isolating the effect of symbolic execution artifacts on deobfuscation performance.
\end{itemize}

We apply this setup to three models introduced in Section~\ref{subsec:models-used}:

\begin{itemize}
  \item \textbf{GPT-4.1-mini-2025-04-14}, a frontier model accessed via API with a context window of approximately 1M tokens.
  \item \textbf{Ministral-8B-latest}, an 8B-parameter open-source model with a 128k-token context window, representing an efficient local or research-friendly deployment.
  \item \textbf{Codestral-latest}, a code-specialized open-source model with a 256k-token context window.
\end{itemize}

Table~\ref{tab:model_context} summarizes the context lengths.
For each model, we evaluate both the fine-tuned variants (baseline vs.\ enhanced) and the corresponding base models without any task-specific training, enabling the comparison in RQ4.

\subsection{Implementation Details}
\label{subsec:implementation}

All symbolic execution runs use KLEE 1.4.0.0 on top of LLVM 3.4, via the Docker environment included in our setup. We run KLEE on the original (unobfuscated) programs to generate SMT2 constraints, KQuery expressions, \texttt{istats} reports, and \texttt{ktest} files.

As described in Section~\ref{subsec:dataset}, we control artifact size purely through dataset filtering rather than truncation. In particular, we first discard programs that trigger path explosion in \textit{KLEE}, and then remove any programs whose combined obfuscated code and KLEE artifacts would exceed the maximum token window of our target LLMs. After this filtering, all remaining instances fit within the model context, so no additional cropping or token-level truncation is required during prompt construction.

Training data are serialized in JSONL format as described in Section~\ref{subsec:training-data-format}, using system messages to define the task, user messages to supply obfuscated code (and KLEE artifacts, when applicable), and assistant messages to provide the desired deobfuscated outputs.

\section{Results}
\label{sec:results}

We now present our empirical findings, organized by research question.
Unless otherwise noted, we report averages over all test programs of a given transformation, model, and training condition.

\subsection{RQ1: Syntactic Correctness}
\label{subsec:rq1-syntax}

\begin{figure}[!htbp]
  \centering
  \includegraphics[width=0.8\textwidth]{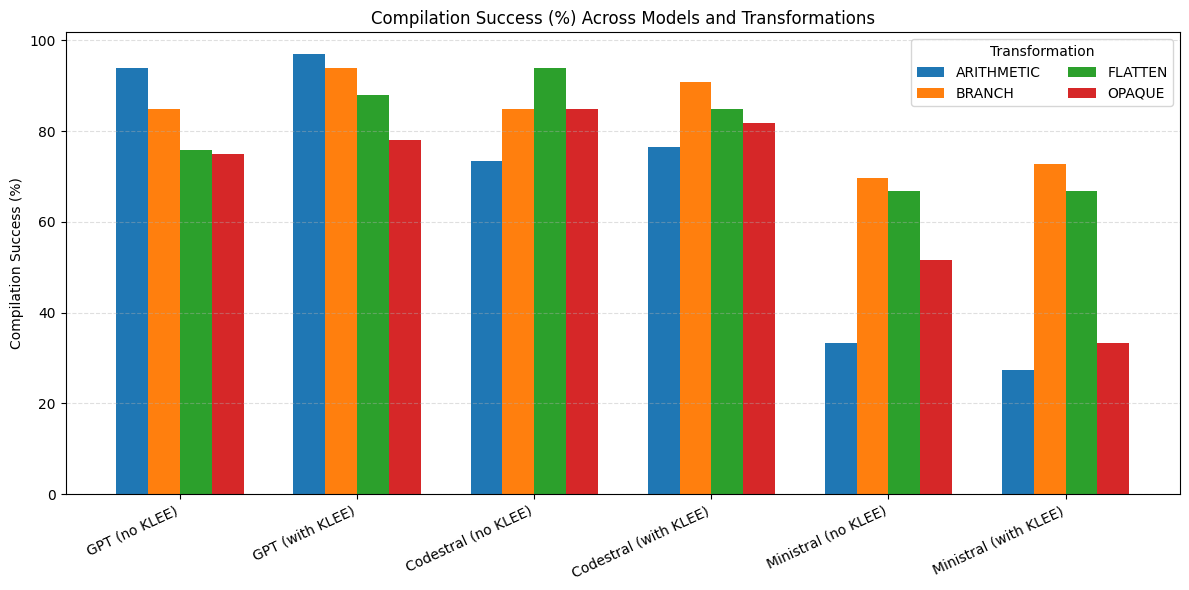}
  \caption{Visualization on Compilation Success (Fine-tuned Model)}
  \label{fig:syntax_finetuned}
\end{figure}

Figure~\ref{fig:syntax_finetuned} summarizes compilation success rates for fine-tuned models.
GPT-4.1-mini is consistently the most reliable: under Arithmetic Encoding, it reaches 96.9\% success with KLEE artifacts versus 93.8\% without.
Similar gains appear for Branch (93.9\% vs.\ 84.8\%) and Flatten (87.9\% vs.\ 75.8\%).
Even for Opaque Predicates, the hardest transformation, GPT-4.1-mini maintains 78.1\% (with KLEE) versus 75.0\% (without).

Codestral also performs strongly, particularly on Branch and Flatten, whereas Ministral lags behind, especially on Arithmetic (27.3\% with KLEE, 33.3\% without) and Opaque (33.3\% vs.\ 51.5\%).

Base-model results in Table~\ref{tab:base_syntax_success} show that task-specific fine-tuning is crucial: GPT and Codestral base models compile reasonably well without KLEE, but often degrade when raw KLEE artifacts are injected without training, indicating that unadapted symbolic data can confuse base models.
Ministral base models perform poorly across the board.

\subsection{RQ2: Semantic Preservation}
\label{subsec:rq2-semantics}

\begin{figure}[!htbp]
  \centering
  \includegraphics[width=0.8\textwidth]{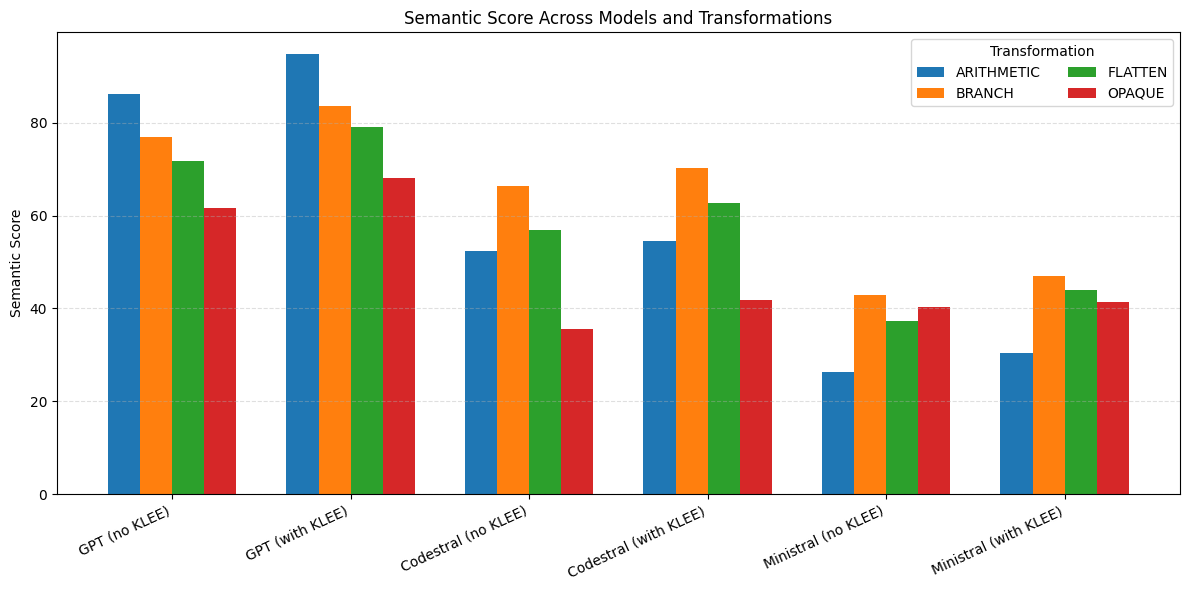}
  \caption{Visualization on Semantic Score (Fine-tuned Model)}
  \label{fig:semantic_finetuned}
\end{figure}

Semantic scores in Tables~\ref{tab:base_semantic_score} and Figure~\ref{fig:semantic_finetuned} tell a similar story.
For fine-tuned GPT-4.1-mini, KLEE artifacts consistently improve semantics: on Arithmetic, the score rises from 86.08 (no KLEE) to 94.70 (with KLEE); improvements of 6--7 points also appear for Branch, Flatten, and Opaque.
Codestral shows smaller but consistent gains (e.g., +2.11 on Arithmetic, +5.67 on Flatten), while Ministral improves modestly yet remains substantially weaker.

For base models, Codestral aligns best with KLEE: on Flatten, its score increases from 52.82 (no KLEE) to 63.33 (with KLEE).
GPT shows partial benefits (e.g., 47.13 to 54.80 on Arithmetic), but gains are limited without fine-tuning.
Ministral base models perform poorly, with some semantic scores close to random behavior.
Overall, these results confirm that symbolic execution artifacts can significantly improve semantic fidelity, but only when the model is trained to interpret them.

\subsection{RQ3: Readability and Quality}
\label{subsec:rq3-quality}

\begin{figure}[!htbp]
  \centering
  \includegraphics[width=0.8\textwidth]{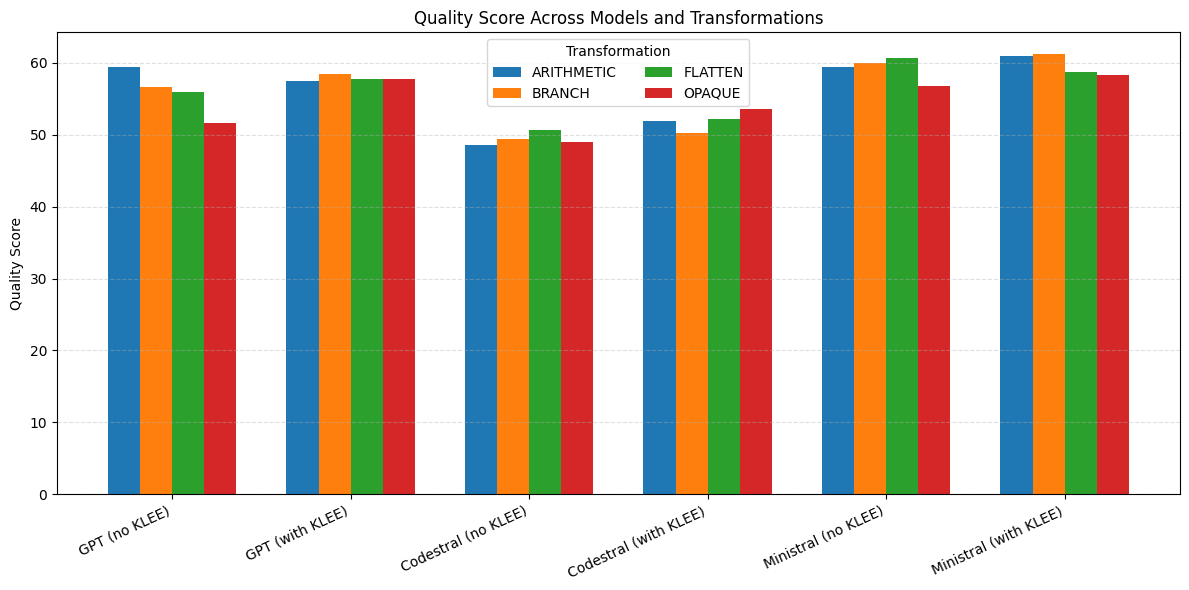}
  \caption{Visualization on Quality Score (Fine-tuned Model)}
  \label{fig:quality_finetuned}
\end{figure}

Quality scores (Tables~\ref{tab:base_quality_score} and Figure~\ref{fig:quality_finetuned}) reveal a different pattern.
Ministral often produces the highest readability scores (around 59--61) among fine-tuned models, despite its weaker syntax and semantics.
GPT-4.1-mini generates slightly less polished code but with far stronger behavioral preservation.

For GPT-4.1-mini, KLEE artifacts do not uniformly improve readability: under Arithmetic, quality slightly decreases (59.38 to 57.45), while Branch and Flatten see modest gains and Opaque shows a larger improvement (51.65 to 57.78).
Codestral displays more consistent quality improvements with KLEE across all transformations.

Base models show relatively stable quality scores (low 60s for GPT and Ministral, high 50s for Codestral), highlighting that readability is not strongly correlated with correctness: models can produce neat, well-formatted code that nonetheless diverges semantically from the original.

\subsection{RQ4: Effect of Fine-tuning and KLEE}
\label{subsec:rq4-comparison}

Comparing base and fine-tuned models across all metrics, we observe:

\begin{itemize}
  \item Fine-tuning drastically improves both compilation and semantics for all models, especially GPT-4.1-mini.
  \item KLEE artifacts are beneficial only when the model is fine-tuned to interpret them; for base models, adding KLEE often \emph{reduces} compilation success and yields mixed semantic gains.
  \item Among fine-tuned models, GPT-4.1-mini benefits the most from KLEE in both syntax and semantics, whereas quality effects are more nuanced.
\end{itemize}

\subsection{RQ5: Transformation Sensitivity}
\label{subsec:rq5-transformations}

Across models and configurations, Opaque Predicates consistently emerge as the most challenging transformation, with the lowest compilation and semantic scores and the largest program sizes.
Arithmetic Encoding and Branch Encoding are generally easier, especially for GPT-4.1-mini and Codestral, while Control-Flow Flattening lies between these extremes.

KLEE artifacts most strongly help in the Opaque regime, where both GPT-4.1-mini and Codestral show sizable semantic and quality gains, indicating that symbolic path information is particularly valuable when control flow is heavily distorted.

\begin{table}[htbp]
\centering

\caption{Compilation Success Rates by Model and Training Type (Fine-tuned Model)}
\label{tab:syntax_success}
\resizebox{0.9\textwidth}{!}{%
\begin{tabular}{|l|l|r|r|r|}
\hline
\textbf{Transformation} & \textbf{Model-Training} & \textbf{Total Files} & \textbf{Successful} & \textbf{Success Rate \%} \\
\hline
ARITHMETIC & CODESTRAL\_withklee & 34 & 26 & 76.5 \\
& CODESTRAL\_noklee & 34 & 25 & 73.5 \\
& GPT4\_withklee & 32 & 31 & 96.9 \\
& GPT4\_noklee & 32 & 30 & 93.8 \\
& MINISTRAL\_withklee & 33 & 9 & 27.3 \\
& MINISTRAL\_noklee & 33 & 11 & 33.3 \\
\hline
BRANCH & CODESTRAL\_withklee & 33 & 30 & 90.9 \\
& CODESTRAL\_noklee & 33 & 28 & 84.8 \\
& GPT4\_withklee & 33 & 31 & 93.9 \\
& GPT4\_noklee & 33 & 28 & 84.8 \\
& MINISTRAL\_withklee & 33 & 24 & 72.7 \\
& MINISTRAL\_noklee & 33 & 23 & 69.7 \\
\hline
FLATTEN & CODESTRAL\_withklee & 33 & 28 & 84.8 \\
& CODESTRAL\_noklee & 33 & 31 & 93.9 \\
& GPT4\_withklee & 33 & 29 & 87.9 \\
& GPT4\_noklee & 33 & 25 & 75.8 \\
& MINISTRAL\_withklee & 33 & 22 & 66.7 \\
& MINISTRAL\_noklee & 33 & 22 & 66.7 \\
\hline
OPAQUE & CODESTRAL\_withklee & 33 & 27 & 81.8 \\
& CODESTRAL\_noklee & 33 & 28 & 84.8 \\
& GPT4\_withklee & 32 & 25 & 78.1 \\
& GPT4\_noklee & 32 & 24 & 75.0 \\
& MINISTRAL\_withklee & 33 & 11 & 33.3 \\
& MINISTRAL\_noklee & 33 & 17 & 51.5 \\
\hline
\end{tabular}
} 
\vspace{1em} 

\begin{minipage}{0.48\textwidth}
\centering
\captionof{table}{Quality Score Results by Model and Training Type (Fine-tuned Model)}
\label{tab:quality_score}
\resizebox{\textwidth}{!}{%
\begin{tabular}{|l|l|r|r|}
\hline
\textbf{Transformation} & \textbf{Model-Training} & \textbf{Std} & \textbf{Score} \\
\hline
ARITHMETIC & CODESTRAL\_withklee & 12.50 & 51.91 \\
& CODESTRAL\_noklee & 11.07 & 48.64 \\
& GPT4\_withklee & 16.42 & 57.45 \\
& GPT4\_noklee & 13.01 & 59.38 \\
& MINISTRAL\_withklee & 11.95 & 60.94 \\
& MINISTRAL\_noklee & 12.97 & 59.48 \\
\hline
BRANCH & CODESTRAL\_withklee & 11.54 & 50.22 \\
& CODESTRAL\_noklee & 12.53 & 49.41 \\
& GPT4\_withklee & 13.02 & 58.45 \\
& GPT4\_noklee & 13.11 & 56.67 \\
& MINISTRAL\_withklee & 14.19 & 61.19 \\
& MINISTRAL\_noklee & 11.88 & 60.00 \\
\hline
FLATTEN & CODESTRAL\_withklee & 11.05 & 52.17 \\
& CODESTRAL\_noklee & 9.79 & 50.72 \\
& GPT4\_withklee & 14.45 & 57.70 \\
& GPT4\_noklee & 16.59 & 55.98 \\
& MINISTRAL\_withklee & 11.52 & 58.75 \\
& MINISTRAL\_noklee & 10.36 & 60.61 \\
\hline
OPAQUE & CODESTRAL\_withklee & 12.65 & 53.57 \\
& CODESTRAL\_noklee & 11.44 & 48.95 \\
& GPT4\_withklee & 13.79 & 57.78 \\
& GPT4\_noklee & 14.80 & 51.65 \\
& MINISTRAL\_withklee & 16.79 & 58.25 \\
& MINISTRAL\_noklee & 17.23 & 56.84 \\
\hline
\end{tabular}
}
\end{minipage}
\hfill
\begin{minipage}{0.48\textwidth}
\centering
\captionof{table}{Semantic Score Results by Model and Training Type (Fine-tuned Model)}
\label{tab:semantic_score}
\resizebox{\textwidth}{!}{%
\begin{tabular}{|l|l|r|r|}
\hline
\textbf{Transformation} & \textbf{Model-Training} & \textbf{Std} & \textbf{Score} \\
\hline
ARITHMETIC & CODESTRAL\_withklee & 45.40 & 54.47 \\
& CODESTRAL\_noklee & 45.89 & 52.36 \\
& GPT4\_withklee & 13.23 & 94.70 \\
& GPT4\_noklee & 30.81 & 86.08 \\
& MINISTRAL\_withklee & 46.65 & 30.34 \\
& MINISTRAL\_noklee & 43.94 & 26.21 \\
\hline
BRANCH & CODESTRAL\_withklee & 44.67 & 70.33 \\
& CODESTRAL\_noklee & 45.47 & 66.45 \\
& GPT4\_withklee & 35.82 & 83.64 \\
& GPT4\_noklee & 39.91 & 76.84 \\
& MINISTRAL\_withklee & 48.94 & 47.07 \\
& MINISTRAL\_noklee & 48.02 & 42.80 \\
\hline
FLATTEN & CODESTRAL\_withklee & 48.35 & 62.62 \\
& CODESTRAL\_noklee & 48.62 & 56.95 \\
& GPT4\_withklee & 39.16 & 78.97 \\
& GPT4\_noklee & 43.74 & 71.82 \\
& MINISTRAL\_withklee & 48.17 & 44.01 \\
& MINISTRAL\_noklee & 47.04 & 37.24 \\
\hline
OPAQUE & CODESTRAL\_withklee & 48.57 & 41.88 \\
& CODESTRAL\_noklee & 47.05 & 35.53 \\
& GPT4\_withklee & 44.21 & 67.98 \\
& GPT4\_noklee & 46.66 & 61.67 \\
& MINISTRAL\_withklee & 46.97 & 41.30 \\
& MINISTRAL\_noklee & 46.93 & 40.27 \\
\hline
\end{tabular}
}
\end{minipage}

\end{table}

\begin{table}[htbp]
\centering

\caption{Compilation Success Rates by Model and Training Type (Base Model)}
\label{tab:base_syntax_success}
\resizebox{0.9\textwidth}{!}{
\begin{tabular}{|l|l|r|r|r|}
\hline
\textbf{Transformation} & \textbf{Model-Training} & \textbf{Total Files} & \textbf{Successful} & \textbf{Success Rate \%} \\
\hline
ARITHMETIC & CODESTRAL\_withklee & 33 & 24 & 72.7 \\
& CODESTRAL\_noklee & 33 & 31 & 93.9 \\
& GPT\_withklee & 33 & 25 & 75.8 \\
& GPT\_noklee & 33 & 31 & 93.9 \\
& MINISTRAL\_withklee & 33 & 8 & 24.2 \\
& MINISTRAL\_noklee & 33 & 5 & 15.2 \\
\hline
BRANCH & CODESTRAL\_withklee & 33 & 29 & 87.9 \\
& CODESTRAL\_noklee & 33 & 28 & 84.8 \\
& GPT\_withklee & 33 & 25 & 75.8 \\
& GPT\_noklee & 33 & 29 & 87.9 \\
& MINISTRAL\_withklee & 33 & 15 & 45.5 \\
& MINISTRAL\_noklee & 33 & 17 & 51.5 \\
\hline
FLATTEN & CODESTRAL\_withklee & 33 & 31 & 93.9 \\
& CODESTRAL\_noklee & 33 & 32 & 97.0 \\
& GPT\_withklee & 33 & 24 & 72.7 \\
& GPT\_noklee & 33 & 30 & 90.9 \\
& MINISTRAL\_withklee & 33 & 16 & 48.5 \\
& MINISTRAL\_noklee & 33 & 12 & 36.4 \\
\hline
OPAQUE & CODESTRAL\_withklee & 33 & 23 & 69.7 \\
& CODESTRAL\_noklee & 33 & 26 & 78.8 \\
& GPT\_withklee & 33 & 24 & 72.7 \\
& GPT\_noklee & 33 & 31 & 93.9 \\
& MINISTRAL\_withklee & 33 & 11 & 33.3 \\
& MINISTRAL\_noklee & 33 & 16 & 48.5 \\
\hline
\end{tabular}
} 

\vspace{1em} 

\begin{minipage}{0.48\textwidth}
\centering
\caption{Quality Score Results by Model and Training Type (Base Model)}
\label{tab:base_quality_score}
\resizebox{\textwidth}{!}{%
\begin{tabular}{|l|l|r|r|}
\hline
\textbf{Transformation} & \textbf{Model-Training} & \textbf{Std} & \textbf{Score} \\
\hline
ARITHMETIC & CODESTRAL\_withklee & 13.38 & 59.97 \\
& CODESTRAL\_noklee & 11.75 & 51.63 \\
& GPT4\_withklee & 10.75 & 67.42 \\
& GPT4\_noklee & 11.94 & 62.48 \\
& MINISTRAL\_withklee & 11.49 & 62.93 \\
& MINISTRAL\_noklee & 12.02 & 60.11 \\
\hline
BRANCH & CODESTRAL\_withklee & 12.71 & 54.29 \\
& CODESTRAL\_noklee & 13.37 & 54.99 \\
& GPT4\_withklee & 10.59 & 65.23 \\
& GPT4\_noklee & 10.43 & 62.25 \\
& MINISTRAL\_withklee & 10.79 & 62.37 \\
& MINISTRAL\_noklee & 10.31 & 61.93 \\
\hline
FLATTEN & CODESTRAL\_withklee & 10.94 & 55.71 \\
& CODESTRAL\_noklee & 10.67 & 53.22 \\
& GPT4\_withklee & 9.28 & 62.60 \\
& GPT4\_noklee & 10.12 & 61.32 \\
& MINISTRAL\_withklee & 11.47 & 63.28 \\
& MINISTRAL\_noklee & 11.03 & 60.92 \\
\hline
OPAQUE & CODESTRAL\_withklee & 13.52 & 55.74 \\
& CODESTRAL\_noklee & 11.69 & 54.05 \\
& GPT4\_withklee & 11.11 & 63.30 \\
& GPT4\_noklee & 11.77 & 61.98 \\
& MINISTRAL\_withklee & 13.72 & 56.70 \\
& MINISTRAL\_noklee & 15.70 & 54.32 \\
\hline

\end{tabular}
}
\end{minipage}
\hfill
\begin{minipage}{0.48\textwidth}
\centering
\caption{Semantic Score Results by Model and Training Type (Base Model)}
\label{tab:base_semantic_score}
\resizebox{\textwidth}{!}{%
\begin{tabular}{|l|l|r|r|}
\hline
\textbf{Transformation} & \textbf{Model-Training} & \textbf{Std} & \textbf{Score} \\
\hline
ARITHMETIC & CODESTRAL\_withklee & 46.46 & 51.17 \\
& CODESTRAL\_noklee & 45.69 & 54.59 \\
& GPT4\_withklee & 46.50 & 54.80 \\
& GPT4\_noklee & 45.59 & 47.13 \\
& MINISTRAL\_withklee & 32.33 & 13.89 \\
& MINISTRAL\_noklee & 16.12 & 3.97 \\
\hline
BRANCH & CODESTRAL\_withklee & 43.41 & 59.90 \\
& CODESTRAL\_noklee & 45.20 & 44.79 \\
& GPT4\_withklee & 45.69 & 52.55 \\
& GPT4\_noklee & 44.14 & 39.86 \\
& MINISTRAL\_withklee & 43.93 & 32.17 \\
& MINISTRAL\_noklee & 41.11 & 25.51 \\
\hline
FLATTEN & CODESTRAL\_withklee & 41.77 & 63.33 \\
& CODESTRAL\_noklee & 43.91 & 52.82 \\
& GPT4\_withklee & 46.14 & 50.19 \\
& GPT4\_noklee & 43.96 & 39.07 \\
& MINISTRAL\_withklee & 39.97 & 23.54 \\
& MINISTRAL\_noklee & 40.11 & 22.40 \\
\hline
OPAQUE & CODESTRAL\_withklee & 46.23 & 51.24 \\
& CODESTRAL\_noklee & 46.30 & 46.34 \\
& GPT4\_withklee & 46.14 & 50.19 \\
& GPT4\_noklee & 45.62 & 52.26 \\
& MINISTRAL\_withklee & 36.83 & 18.48 \\
& MINISTRAL\_noklee & 44.36 & 30.80 \\
\hline

\end{tabular}
}
\end{minipage}

\end{table}

\begin{figure}[!htbp]
  \centering
  \includegraphics[width=0.8\textwidth]{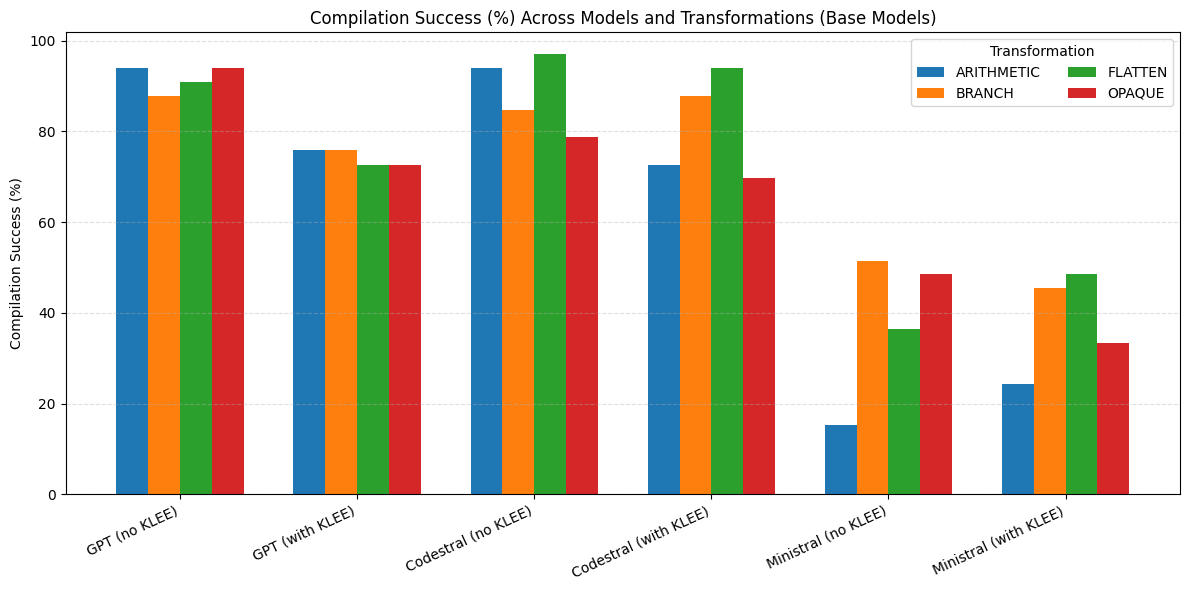}
  \caption{Visualization on Compilation Success (Base Model)}
  \label{fig:basemodel_compile}
\end{figure}

\begin{figure}[!htbp]
  \centering
  \includegraphics[width=0.8\textwidth]{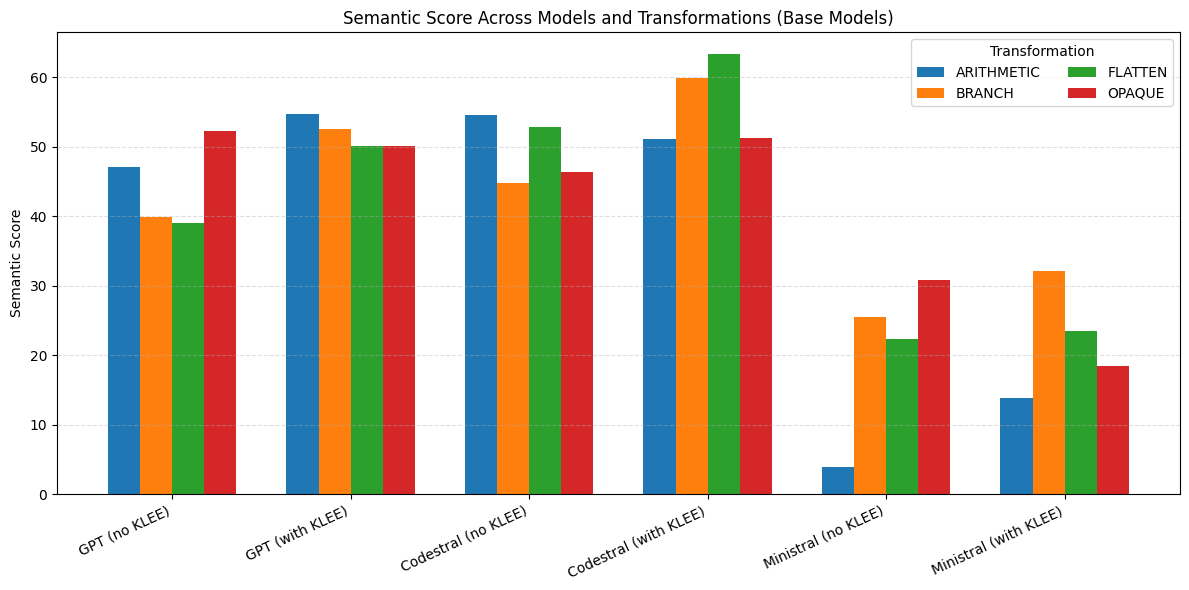}
  \caption{Visualization on Semantic Score (Base Model)}
  \label{fig:basemodel_semantic}
\end{figure}

\begin{figure}[!htbp]
  \centering
  \includegraphics[width=0.8\textwidth]{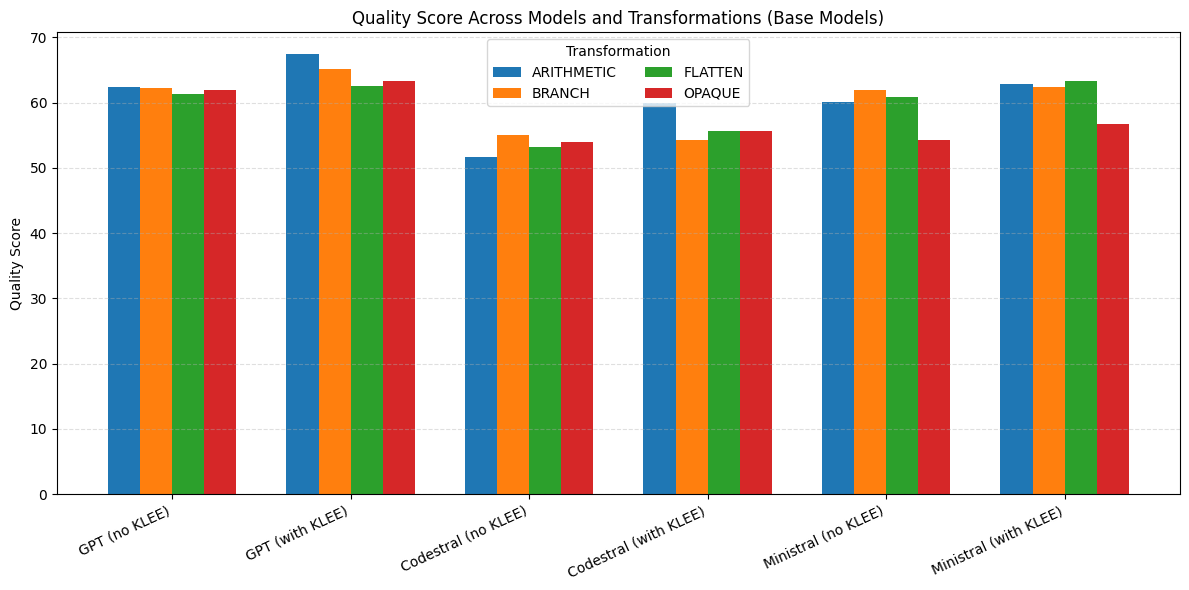}
  \caption{Visualization on Quality Score (Base Model)}
  \label{fig:basemodel_quality}
\end{figure}
\section{Related Works}

\subsection{Obfuscation and Deobfuscation Techniques}

Obfuscation has long been studied as both a defensive and offensive technique in software. Its primary goal is to transform code so that it becomes harder to understand, reverse engineer, or analyze, while still preserving the original semantics. Surveys such as Schrittwieser et al.\ and Xu et al.\ highlight how obfuscation is widely used in industry to protect intellectual property, but also emphasize that it complicates essential software engineering tasks such as program comprehension, debugging, and maintenance~\cite{schrittwieser2016protecting,xu2017secure}.

Many practical obfuscation transformations have been proposed. Classic techniques include control-flow flattening, opaque predicates, and virtualization. Control-flow flattening has been adapted to C++ programs~\cite{laszlo2009obfuscating}, and even formally verified for semantics preservation in CompCert~\cite{blazy2016formal}. Opaque predicates, which insert always-true or always-false conditions to confuse analysis, have been extended to generalized dynamic forms that can embed into loops and branches~\cite{xu2016generalized}. Recent work has also proposed heuristics and machine learning approaches to detect or neutralize opaque predicates~\cite{tung2020heuristic,tofighi2019defeating}. On the virtualization side, researchers have explored symbolic execution combined with compilation optimization to recover semantics from virtualized code~\cite{salwan2018symbolic,liang2017deobfuscation}.

Tool support has made these techniques widely accessible. \textit{Tigress}, a source-to-source obfuscator for C, provides a wide range of transformations and has been used in both teaching and research~\cite{taylor2016tool}. \textit{Obfuscator-LLVM} (OLLVM) introduced transformations such as bogus control-flow and instruction substitution directly in LLVM, and has since become a standard baseline in many experimental studies~\cite{junod2015obfuscator}. These tools have enabled systematic comparisons of obfuscation strength and resilience~\cite{sebastian2017predicting,ollivier2019obfuscation}, and even pushed forward new evaluation frameworks that suggest moving beyond traditional potency and resilience metrics~\cite{de2025new}.

Despite this progress, most of the research in this space has focused on how to design and evaluate obfuscations, rather than how to reliably deobfuscate code at scale. For software engineering tasks such as automated testing, bug detection, or program comprehension, effective deobfuscation is just as important. Existing deobfuscation attempts, whether based on symbolic execution, heuristics, or learning, often break down under complex or chained transformations. This gap motivates the need for new approaches that can combine the strengths of different techniques to restore analyzable code.

\subsection{Symbolic Execution for Analysis and Deobfuscation}

Symbolic execution has been a core technique in program analysis for decades. Surveys of the field provide a broad view of its strengths and challenges, especially issues like path explosion and constraint solving~\cite{baldoni2018survey,duraibi2019survey}. Over the years, several practical systems have shown that symbolic execution can achieve high coverage and uncover deep bugs. A well-known example is \textit{KLEE}, which automatically generated high-coverage tests for GNU coreutils and exposed numerous previously unknown errors~\cite{cadar2008klee}. Other frameworks extended this idea to broader contexts, such as \textit{S2E}, which allowed selective symbolic execution of specific program parts while concretely executing the rest~\cite{chipounov2009selective}, and \textit{SymDrive}, which applied symbolic execution to driver testing without hardware~\cite{renzelmann2012symdrive}. These works underline the adaptability of symbolic execution across many software engineering domains.

When it comes to deobfuscation, symbolic execution has also been explored as a way to recover hidden semantics. Salwan et al.\ proposed symbolic deobfuscation from virtualized code back to original semantics~\cite{salwan2018symbolic}, while Liang et al.\ combined symbolic execution with compiler optimizations to deobfuscate virtualization-based protections~\cite{liang2017deobfuscation}. Other research has focused on resilience prediction: Banescu et al.\ built machine learning models to predict how long specific obfuscations can resist symbolic execution~\cite{sebastian2017predicting}. On the defensive side, Ollivier et al.\ surveyed anti-dynamic symbolic execution (anti-DSE) protections, showing that while many techniques exist, symbolic execution is still highly effective against most real-world obfuscations~\cite{ollivier2019obfuscation}.

Despite its strengths, symbolic execution by itself often struggles when facing modern obfuscations. Path explosion makes complex transformations intractable, and solver limitations mean some opaque predicates or flattened control flows remain unresolved~\cite{xu2016generalized,tung2020heuristic}. As a result, symbolic execution is valuable but not sufficient on its own for deobfuscation in large-scale or heavily protected programs. This motivates approaches that combine symbolic reasoning with other techniques, such as machine learning or large language models, to achieve more reliable results.

\subsection{Machine Learning for Deobfuscation and Reverse Engineering}

Before the rise of large language models, researchers explored machine learning as a way to support reverse engineering and deobfuscation. One active area was variable and symbol recovery from stripped binaries. For example, Banerjee et al.\ applied constrained masked language modeling to predict variable names from decompiled binaries~\cite{banerjee2021variable}, while He et al.\ introduced \textit{Debin} to recover debug information such as names and types~\cite{he2018debin}. More recently, \textit{ReSym} combined LLM components with symbolic reasoning to improve recovery of variables and data structures~\cite{xie2024resym}. These approaches highlight how even lightweight learning models can restore meaningful semantics that are lost through compilation or obfuscation.

Machine learning has also been applied to more complex problems like Android deobfuscation and malicious code analysis. Su et al.\ proposed \textit{MACNETO}, a deep learning system that matched obfuscated Android applications to their original versions with high precision~\cite{su2017deobfuscating}, while Kan et al.\ automated the deobfuscation of Android native code by combining taint analysis and symbolic execution~\cite{kan2019automated}. More recently, Patsakis et al.\ evaluated state-of-the-art LLMs on real malware campaigns, showing that while these models can interpret obfuscated scripts, fine-tuning is essential for reliability~\cite{patsakis2024assessing}.

Another important direction was the use of neural decompilers. Fu et al.\ introduced \textit{Coda}, one of the first neural program decompilers that significantly outperformed rule-based approaches~\cite{fu2019coda}. Later works extended this line to optimized binaries~\cite{kan2019automated} and retargetable decompilation using neural translation~\cite{hosseini2022beyond}. These efforts revealed the potential of neural models to handle code transformation tasks that traditional analysis tools struggled with.

While these machine learning methods showed promise, they also had limitations. Most systems focused on narrow tasks, such as variable renaming or deobfuscating a specific platform, and lacked generalizability across different obfuscation transformations. Moreover, they rarely provided guarantees of correctness or semantic fidelity. These shortcomings set the stage for the introduction of large language models, which offered broader capabilities but still faced their own challenges.

\subsection{Large Language Models for Program Analysis and Deobfuscation}

With the rapid progress of large language models (LLMs), researchers have begun to apply them directly to program analysis tasks. Several works target variable and function name recovery, where LLMs have proven effective at capturing semantic context. Xu et al.\ used generative modeling to recover variable names from stripped binaries, showing improvements over classification-based baselines~\cite{xu2025unleashing}. Jiang et al.\ extended this idea with \textit{SYMGEN}, a domain-adapted LLM that infers function names in stripped binaries, achieving large gains over earlier models~\cite{jiang2025beyond}. Other work like \textit{DIRECT} applied transformers for variable name recovery from decompiled binaries~\cite{nitin2021direct}. These studies illustrate how LLMs can restore lost semantic information in ways earlier ML approaches struggled with.

Beyond symbol recovery, LLMs have been used for full decompilation and deobfuscation. Tan et al.\ introduced \textit{LLM4Decompile}, an open-source effort to train LLMs specifically for decompiling binaries into high-level code~\cite{tan2024llm4decompile}. Hu et al.\ proposed \textit{DeGPT}, which improves decompiler output readability and semantic clarity by coordinating multiple LLM roles~\cite{hu2024degpt}. Zou et al.\ refined this direction with \textit{D-LiFT}, a fine-tuning framework that optimizes code quality metrics and integrates verification signals~\cite{zou2025d}. Other approaches like \textit{Idioms} jointly predict code and type definitions to improve neural decompilation~\cite{dramko2025idioms}. Together, these systems show that LLMs can be integrated as both frontends and backends in decompilation pipelines.

At the same time, several studies have benchmarked LLMs directly on deobfuscation tasks. Chen et al.\ introduced \textit{JsDeObsBench} to evaluate LLMs on JavaScript deobfuscation, finding that models simplify code but still struggle with syntax reliability~\cite{chen2025jsdeobsbench}. Beste et al.\ explored LLMs on five different obfuscation types, demonstrating that fine-tuned models reduce code complexity but still fail in chained settings~\cite{beste2025exploring}. Tkachenko et al.\ built a four-dimensional framework for evaluating LLMs on assembly-level obfuscations, and showed systematic weaknesses like predicate misinterpretation and context integration errors~\cite{tkachenko2025deconstructing}. These benchmarks highlight both progress and persistent challenges.

There is also growing interest in hybrid and agentic frameworks. Ching et al.\ proposed \textit{ALFREDO}, an agent-based LLM framework for deobfuscation that integrates external tools to guide iterative refinement~\cite{nataliealfredo}. Jiang et al.\ presented \textit{CASCADE}, a production-ready LLM-powered JavaScript deobfuscator deployed at Google, which combines Gemini with intermediate representations to restore obfuscated APIs~\cite{jiang2025cascade}. Liang et al.\ introduced \textit{DeCoda}, which combines graph learning with LLM-based recovery to improve detection of malicious JavaScript~\cite{liang2025breaking}. These works suggest that LLMs benefit from being part of larger systems, rather than being used in isolation.

Despite these advances, a consistent limitation across LLM-based studies is reliability. Models frequently generate code that does not compile, fails to preserve semantics, or hallucinates constructs~\cite{nikiema2025code,tkachenko2025deconstructing}. While surveys of LLMs for software security and source code analysis~\cite{jelodar2025large,jelodar2025largedataset} show the breadth of applications, they also underscore that deobfuscation remains an open challenge. In particular, no prior work has systematically combined symbolic execution artifacts with LLMs to improve compilation success and semantic preservation. This gap is what our study aims to address.

\subsection{Our Contribution in Context}

Across these strands of research, we see three patterns. First, the obfuscation community has created a wide variety of transformations and tools, but most work emphasizes how to make obfuscation stronger rather than how to systematically reverse it~\cite{schrittwieser2016protecting,junod2015obfuscator}. Second, symbolic execution has shown strong potential for revealing hidden program behavior, yet it often collapses under complex or chained obfuscations because of scalability limits~\cite{sebastian2017predicting,ollivier2019obfuscation}. Third, LLM-based approaches have opened new opportunities in program analysis, but they remain unreliable, producing code that does not always compile or preserve semantics~\cite{chen2025jsdeobsbench,tkachenko2025deconstructing}.

What has not been fully explored is how these two directions—symbolic execution and LLMs—can be combined in a complementary way. Our work addresses this gap by fine-tuning LLMs with symbolic artifacts generated by \textit{KLEE}. The intuition is that symbolic execution provides hard semantic facts, while LLMs bring flexible reasoning and code synthesis. By bringing them together, we obtain models that produce deobfuscated code which is more compilable, semantically faithful, and readable than either method alone. To the best of our knowledge, this is the first study to systematically evaluate this hybrid approach.

\bibliographystyle{ACM-Reference-Format}
\bibliography{main}

\end{document}